\documentstyle[aps,prl,psfig,twocolumn]{revtex}
 \def\ksim{\mathrel{\rlap{\lower0.2em\hbox{$\sim$}}\raise0.2em\hbox{$<$}}}
 \def\gsim{\mathrel{\rlap{\lower0.2em\hbox{$\sim$}}\raise0.2em\hbox{$>$}}}
\begin{document}

\title{
Stringent limits on quark star masses due to the
chiral transition temperature
}
\author{
{\sc A. Peshier$^a$, B. K\"ampfer$^b$, G. Soff$^c$}}
\address{
$^a$ Department of Physics, Brookhaven National Laboratory, 
     Upton, NY 11973, USA\\
$^b$ Forschungszentrum Rossendorf, PF 510119, 01314 Dresden, Germany\\
$^c$ Institut f\"ur Theoretische Physik, TU Dresden, 01062 Dresden, Germany}

\maketitle

\begin{abstract}
Recent finite-temperature QCD lattice data are analyzed within a
quasiparticle model, and extrapolated to nonzero chemical potential.
Determined by the chiral transition temperature, the resulting 
equation of state of charge neutral, $\beta$-stable deconfined matter 
limits the mass and the size of quark stars.
\\[1mm]
{\it PACS number(s):\/}
24.85.+p, 12.38.Mh, 21.65.+f, 26.60.+c
\end{abstract}

\vspace*{3mm}

Soon after the discovery of quarks and gluons as the constituents of
hadrons, the possible existence of quark cores in neutron stars has 
been conjectured \cite{QCD_q_stars}.
Even before establishing quantum chromodynamics (QCD) as fundamental 
theory of strong interaction, the notion of quark matter in stars had 
been introduced by various authors \cite{old_q_stars}.
Since that time many investigations have been performed to elucidate 
various consequences of this hypothesis. 
As one of the most fascinating conjectures, the existence of self-bound 
quark matter \cite{strangelets} and pure quark stars (as surveyed in 
\cite{pure_q_stars}) was proposed. 
Other investigations are devoted to rapidly rotating quark stars 
\cite{Glendenning_Weber}, the cooling behavior of neutron stars with 
quark cores, and the mass-radius relation which is now becoming 
accessible to observation.
If star matter exhibits a strongly first-order phase transition above 
nuclear density, a third stable family \cite{third_island} of  
ultra-dense, cold stars can appear beyond the branches of white dwarfs and 
ordinary neutron stars, thus creating so-called twin stars.
The necessary discontinuity in the equation of state (EoS) may be realized
in nature at the chiral transition \cite{Glendenning_Kettner}, but also 
other phase changes of nuclear matter may allow for 
twin stars \cite{Hanauske}.
The transition of a neutron star to the denser configuration, triggered 
e.g.\ by accretion, can be accompanied by an ejection of the outermost 
layers \cite{our_hydro}. This may result in specific supernova events 
\cite{Hanauske,SN1987A}.

While different scenarios for these phenomena have been explored, most 
studies suffer from uncertainties of the EoS of strongly interacting matter.
Predictions of the hadronic EoS, which can be adjusted to properties of 
nuclear matter, become increasingly model dependent when extrapolated 
above the nuclear saturation density.
For deconfined matter, which we will consider in the following, the
bag model EoS is often applied \cite{QCD_q_stars}, which is, however,
in conflict with thermodynamic QCD lattice data.
Improved approaches take into account medium modifications of the quark 
masses \cite{Schertler} or rely on perturbation theory \cite{Fraga}, but 
also introduce {\it a priori} undetermined model parameters.

Here we are going to describe deconfined matter within a quasiparticle 
picture \cite{PRD_1996,PRC_2000}. 
The approach is tested in the nonperturbative regime by comparison 
to finite-temperature QCD lattice data, which completely fixes the 
intrinsic parameters.
The model yields in a straightforward way the EoS at finite chemical 
potential and vanishing temperature, which is then employed to calculate 
bulk properties of quark stars.

Our starting point is the thermodynamic potential in the form 
$p(T, \mu) = \sum_i p_i - B(\Pi_j^*)$, where the contribution 
$p_i(T, \mu_i(\mu); m_i^2)$ of the thermodynamically relevant partons 
is given by the pressure of an ideal Bose or Fermi gas of quasiparticles 
with mass squared $m_i^2 = m_{0,i}^2+\Pi_i^*$.
$m_{0,i}$ is the rest mass of parton species $i$, and $\Pi_i^*$ its 
asymptotic self-energy at the light cone (depending on the temperature 
$T$ and the chemical potential $\mu$).
Perturbatively, this ansatz is motivated by the fact that in the interacting
system the position of the poles of the transverse gluon propagator and the
quark particle-excitation is, for the dominating hard momenta of the order 
$T, \, \mu$, described approximately by mass shells.
The longitudinal gluon modes and the quark hole-excitations, on the other
hand, are overdamped in this region of the phase space.
The function $B(\Pi_j^*)$ is determined from a thermodynamical
self-consistency condition, via $\partial B/\partial \Pi_j^* =
\partial p_j(T,\mu_j; m_j^2)/\partial m_j^2$.

This phenomenological approach reproduces the leading order perturbative
result for the QCD pressure \cite{PRC_2000}. 
Since it partly resums terms of higher order, it can be supposed to be 
applicable at larger coupling strength.
(This assertion was also made in Refs.~\cite{HTLpt} where the pressure 
or the entropy have been approximated by resumming the hard thermal loop 
self-energies.)
Indeed, parameterizing the effective coupling by
\begin{equation}
 G^2(T, \mu=0)
 =
 \frac{48\pi^2}
  {(33-2N_f)\ln\left(
     \displaystyle\frac{T+T_s}{T_c/\lambda}\right)^{\!\!2}} \, ,
 \label{eq:G(T,0)}
\end{equation}
finite-temperature QCD lattice data can be reproduced quantitatively, 
even in the close vicinity of the chiral transition temperature $T_c$, by 
adjusting $T_s$ and $\lambda$ as demonstrated in \cite{PRD_1996,PRC_2000}
for $N_f = 2$ and 4 light flavors, and in the quenched limit.
Based on that, only estimates could be provided for the relevant case of 
three quark flavors with physical masses.
Now, the predictions of the quasiparticle model can be considerably
refined by systematically analyzing the recent lattice calculations 
\cite{Karsch_EoS} with $N_f=2,3$ and $N_f=2+1$.
First of all, we emphasize the fact that the pressure, as a function 
of $T/T_c$ and scaled by the free value, displays a striking universal 
behavior for different numbers of flavors \cite{Karsch_QM2001}. 
This observation supports the proposed quasiparticle description.

\begin{figure}[t]
\centerline{{\psfig{file=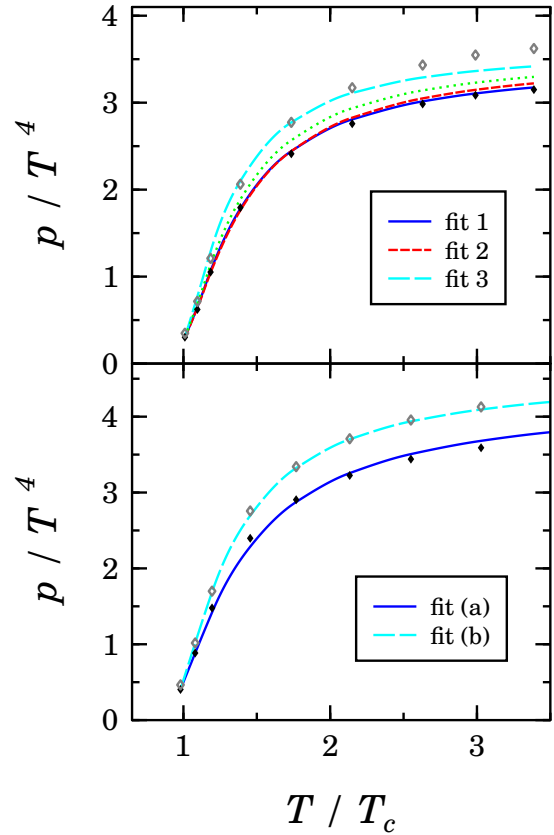,width=8cm,angle=0}}}
\caption{The pressure for $N_f=2$ (upper panel) and 2+1 (lower panel)
   at $\mu=0$.
   The full symbols represent the lattice data \protect\cite{Karsch_EoS},
   the open symbols are estimates for the continuum extrapolation.
   The quasiparticle fits and the dotted curve are explained in the text.
   The parameters $[\lambda,-T_s/T_c]$ for $N_f=2$ are according to
   Table 1, and 
   [3.5, 0.60] for fit (a) and 
   [6.6, 0.78] for fit (b), respectively.
   \label{fig1}}
\end{figure}

In Ref.~\cite{Karsch_EoS}, the pressure $p(T)$ has been calculated 
on the lattice for two and three flavors with a quark (rest) mass 
$m^{\rm latt}=0.4\,T$.
These data can be reproduced by the quasiparticle model assuming either
$m_{0,q}=0.4\,T$ (fit 1 in Fig.~\ref{fig1}a for $N_f=2$) or $m_{0,q}=0$
(fit 2); it naturally appears that the first option yields a slightly
better fit.
To estimate the effect of the temperature dependent quark mass 
$m^{\rm latt}$, we also calculated the pressure within the quasiparticle 
model assuming $m_{0,q}=0$ and with $\lambda$ and $T_s$ from fit 1 
(dotted line in Fig.~\ref{fig1}a).
The resulting variation of the order of 5\% is within the uncertainty 
of the lattice data.
To extrapolate the lattice data to the chiral limit, and to correct for
finite size effects, it has been suggested in \cite{Karsch_EoS} to scale, 
for $2T_c \ksim T \ksim 4T_c$, the `raw' data by a factor of 1.15.
Assuming, as a conservative estimate, the validity of this extrapolation
also for $T_c \ksim T \ksim 2T_c$ (as relevant below), it can be described 
in that interval with the parameters of fit 3. For larger temperatures, 
this fit is still within the estimated errors.
The data for $N_f=3$ are analyzed correspondingly, and the fit parameters 
are summarized in Table 1.

Encouraged by the convincing agreement of the quasiparticle model with the 
finite-temperature lattice data in the nonperturbative regime, it can be 
further extended to nonzero chemical potential
which is not yet accessible to lattice calculations. 
This only assumes that the underlying 
quasiparticle structure does not change\footnote{
   For the equation of state, the phenomenon of color superconductivity 
   plays a less important role since it occurs on the Fermi surface.
   The effect is of the order $(\Delta/\mu)^2$, where the gap $\Delta$ is
   small compared to $\mu$ \cite{Krishna}.}.
As shown in \cite{PRC_2000}, Maxwell's relation imposes a partial 
differential flow equation for the effective coupling $G^2(T,\mu)$, which 
allows to calculate the pressure in parts of the $T, \, \mu$ plane if the 
coupling is known at finite $T$ and $\mu = 0$.
From the pressure $p$, the entropy and particle densities $s$ and $n$ 
follow by differentiating with respect to $T$ and $\mu$, respectively, 
and the energy density is obtained by $e = sT+n\mu-p$.

The above quasiparticle fits in which $m_{0,q}$ do not depend on $T$ 
are mapped to finite quark chemical potential $\mu_q=\mu$ and to zero
temperature.
The difference of 15\% between the fits of the `raw' lattice data and 
the estimated continuum extrapolation is also apparent at $\mu \not= 0$.
Remarkably, however, the relation $e(p)$ at $T=0$ turns out to be rather
similar for both fits. 
In particular, in both cases the energy densities at vanishing pressure 
are approximately equal. 
For $p \ksim 15\,T_c^4$, both curves can be parameterized within 2\% 
accuracy by a linear function
\noindent
\begin{equation}
  e(p) = 4\tilde B + \alpha p \, .
  \label{e(p)}
\end{equation}
The parameter $\tilde B$ is related to the bag constant in the bag model 
EoS, while a slope parameter $\alpha$ different from $\alpha_{\rm bag}=3$
indicates nontrivial interaction effects.
For the $N_f=2$ case we find $ 4\tilde B_{[2]} \approx 19\, T_c^4$ and
$\alpha_{[2]} \approx 3.4 \ldots 3.7$ (the larger value describes fit 2, 
the smaller one fit 3).
For $N_f=3$ we obtain $ 4\tilde B_{[3]} \approx 15 \ldots 16.5\, T_c^4$ and
$ \alpha_{[3]} \approx 3.4 \ldots 3.8$.
The parameters are approximately equal for $N_f=2$ and $N_f=3$, which is, 
given the aforementioned universality of the scaled pressure for different 
numbers of flavors, not unexpected.
Nonetheless, we emphasize again that in both cases the EoS at $T=0$ in the 
form (\ref{e(p)}) is to a noteworthy degree insensitive (in contrast to the 
model parameters $\lambda$ and $T_s$) to the remaining uncertainties of the 
lattice data.

With these remarks in mind, we turn to the discussion of the 2+1 flavor 
case.
The lattice calculation \cite{Karsch_EoS} assumed $m_l^{\rm latt}=0.4\,T$
for the light flavors, and $m_h^{\rm latt}=T$ for the heavy one.
As shown in fit (a) in Fig.~\ref{fig1}b, the data can be described in the
quasiparticle model assuming the same temperature dependent quark rest
masses as on the lattice.
For an estimate of the continuum result and in order to extrapolate to 
physical quark masses, we scale the lattice data again by a factor of 1.15
(motivated by the universality of the pressure for different flavor numbers 
and by the observed weak dependence on the quark rest masses).
The scaled data can be reproduced by the quasiparticle model assuming
$m_{0,l}=0$ for the light flavors $l=u,d$, and $m_{0,s} = T_c$ for the 
strange flavor, see fit (b) in Fig~\ref{fig1}b.

\begin{figure}[hb]
~\vskip-7mm
\centerline{{\psfig{file=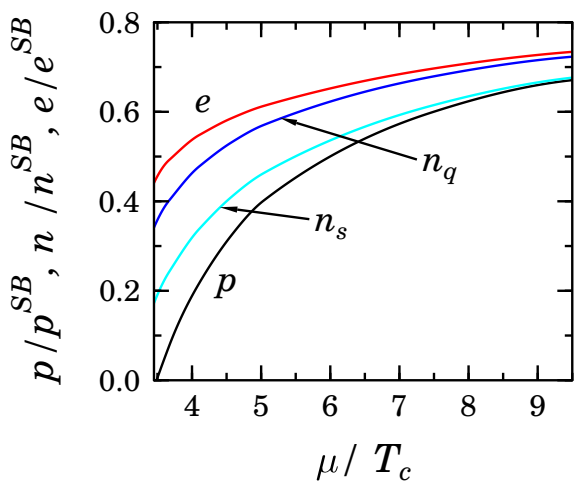,width=8cm,angle=0}}}
 \caption{The pressure, the strange and the net quark density, and the
          energy density for neutral and $\beta$-stable deconfined matter
          at $T=0$, scaled by the Stefan-Boltzmann limits, where all 
          masses can be neglected and $\mu_e$ vanishes.
   \label{fig5}}
\end{figure}

We now apply our model, with the parameters of fit (b), for the case of 
neutral quark matter in $\beta$ equilibrium with leptons. 
The contribution of the leptons is approximated by a free electron gas, so 
the chemical potentials are related by $\mu = \mu_{d,s} = \mu_u + \mu_e$. 
The electron chemical potential $\mu_e(\mu)$ is determined by the 
requirement of electric neutrality. 
The pressure, as well as the quark and energy densities at $T=0$, are 
depicted in Fig.~\ref{fig5}. 
(With respect to \cite{Krishna_1} we mention that
these results are not sensitive to the lepton component.)

The pressure of the deconfined phase vanishes at a chemical potential
$\mu_c \approx 3.5\,T_c$, which provides a lower bound for the transition
to the chirally broken phase.
Lattice calculations \cite{Karsch_T_c} determine $T_c \approx 150 \ldots 
170\,$MeV. The uncertainty has, apart from the scaling, only a marginal 
effect on the thermodynamic properties shown in Fig.~\ref{fig5}.
Assuming $T_c = 160\,$MeV in the following, the net quark density $n_q$ at
$\mu_c = 556\,$MeV is approximately five times the quark density $\rho_0
= 3 \cdot 0.17/$fm$^3$ in nuclear matter.
This corresponds to an energy per baryonic degree of freedom of about 1.6 GeV. 
Consequently, strange quark matter is not stable according to our model.
On the other hand, the value of $n_q(\mu_c)$ imposes a restriction on the EoS 
of hadronic matter, which still exhibits large
theoretical uncertainties for higher densities.
\begin{figure}[ht]
~\vskip-7mm
\centerline{{\psfig{file=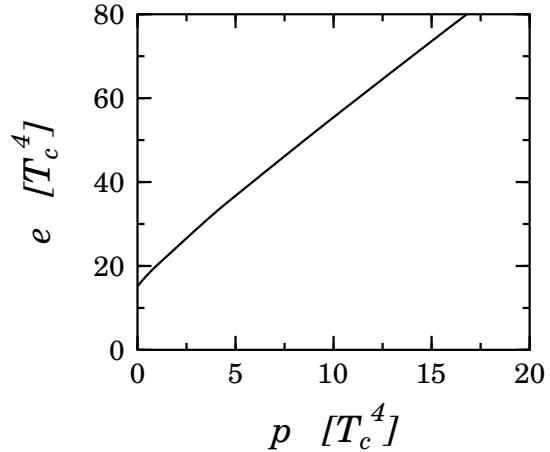,width=7.5cm,angle=0}}}
 \caption{The relation $e(p)$ from Fig.~2.
   \label{fig6}}
\end{figure}
The relation $e(p)$ can again be parameterized by the
linear function
(\ref{e(p)}) (see Fig.~\ref{fig6}) with $\alpha \approx 3.8$ and 
$\tilde{B}^{1/4} \approx 228\,$MeV, which is considerably larger than 
the typical value of $145\,$MeV often used in the bag model.

It is interesting to compare our results to those of the perturbative
approach \cite{Fraga}, where also an almost linear relation $e(p)$ was 
found. For the choice of the renormalization scale $\bar\Lambda = 2 \mu$, 
an  effective bag constant $\approx 210\,$MeV (at $\alpha = 3$ fixed) 
and similar quark densities were obtained despite of a somewhat smaller 
value of $\mu_c$.

The structure of nonrotating stars is determined by the 
Tolman-Oppenheimer-Volkov (TOV) equation (cf.\ \cite{Glendenning_Weber}), 
in which the EoS of the star matter enters only in the form $e(p)$.
In the following, we use the above results and consider stars composed 
entirely of neutral $\beta$-stable deconfined matter. 
We first recall a property of the TOV equation: If $e$ and $p$ are scaled 
by a factor of $\kappa$, the mass and length scales change by a factor of 
$\kappa^{-1/2}$.
With the EoS (2), therefore, the mass and radius of the star are $\propto 
\tilde{B}^{-1/2}$ \cite{strangelets}.
Increasing values of $\alpha$ lead to a softening of the EoS,
which results in a larger pressure gradient in the star and to more 
compact configurations.

\begin{figure}[t]
~\vskip-7mm
\centerline{{\psfig{file=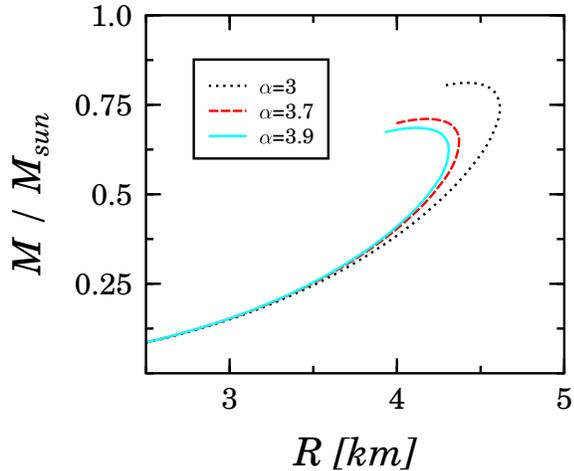,width=8cm,angle=0}}}
 \caption{The mass-radius relation of quark stars for the EoS (2) 
   with $\tilde B^{1/4} = 228\,$MeV and parameters $\alpha$ 
   around the fitted value. Left to the maximum mass, the upper branches 
   become unstable.
   \label{fig7}}
\end{figure}

The mass-radius relation according to our EoS is shown in Fig.~\ref{fig7}.
Compared to results based on the conventional bag model EoS, which 
predict a maximum mass up to $1.5\,M_{\rm sun}$ and a corresponding 
radius $R\approx 10\,$km similar to the bulk properties of neutron stars, 
we find configurations which are smaller and lighter by, approximately, 
a factor of two (rotational effects may increase the maximum mass up to 
30\% \cite{Glendenning_Weber}).
This general statement is robust in our approach.
The main effect is due to the large value of $\tilde B$, whereas the
change $\alpha = 3 \rightarrow 3.8$ reduces the maximum mass and radius 
only by about 15\%.

As pointed out in Refs.~\cite{third_island,Hanauske,Fraga}, an EoS 
allowing for a strong first-order transition to hadronic matter, as 
ours, can lead to a second stable branch in the mass-radius relation
nearby the neutron star peak, thus permitting twin stars.
These features, however, are sensitive to 
\noindent
details of the hadronic EoS 
and of the phase transition and deserve separate investigations.

In summary, our quasiparticle analysis of recent finite-temperature QCD 
lattice data \cite{Karsch_EoS} allows to predict the EoS at nonzero 
chemical potential.
At $T=0$, the relation $e(p)$ is found to be almost linear and insensitive
to the remaining uncertainties of the lattice data.
For QCD with physical quark masses, the resulting energy densities are 
large and exclude self-bound quark nuggets or strangelets.
We find stringent limits on the mass and the size of strange quark stars:
Assuming $T_c = 160\,$MeV, the estimated maximum mass is $M_{\rm max} 
\approx 0.7\, M_{\rm sun}$, with a corresponding radius of $R_{\rm max} 
< 5\,$km; these values scale as $T_c^{-2}$.
Such small but dense objects are of interest with respect to MACHO 
events \cite{Fraga,MACHO}.

Useful discussions with
D.\ Blaschke, J.\ Engels, E.\ Fraga, E.\ Grosse, F.\ Karsch, R.\ Pisarski,
J.\ Schaffner-Bielich, F.\ Thielemann, and F.\ Weber are gratefully 
acknowledged.
The work is supported by the grant BMBF 06DR921.
A.P.\ is supported by DOE under grant DE-AC02-98CH10886 and by the 
A.-v.-Humboldt foundation (Feodor-Lynen program).

\begin{table}
\begin{center}
\begin{tabular}{c|c|c|c}
        &    fit 1    &     fit 2   &    fit 3      \\ \hline
$N_f=2$ & [4.4, 0.73] & [6.0, 0.80] & [11.5, 0.90]  \\
$N_f=3$ & [5.3, 0.73] & [7.8, 0.80] & [15.2, 0.88]
\end{tabular}
\end{center}
\caption{The parameters $[\lambda, -T_s/T_c]$ in the effective coupling
  (\protect\ref{eq:G(T,0)}), fitted to the lattice data 
  \protect\cite{Karsch_EoS}.
  \label{Table1}}
\end{table}
\end{document}